\documentclass[preprint,12pt]{revtex4}
 
\usepackage{revsymb}
\usepackage{graphicx} 
\usepackage{amsmath}
  \usepackage{tikz}

\usepackage{ulem}
\usepackage{setspace}

\DeclareGraphicsExtensions{.jpg,.png}

\begin{document}

\title{The effect of colored noise on heteroclinic orbits}

\author{Jean-R\'{e}gis Angilella}
\address{Universit\'e de Caen Normandie, ESIX Cherbourg, ABTE EA4651-ToxEMAC, 14000 Caen, France}

\begin{abstract}
{
The dynamics of a weakly dissipative Hamiltonian system submitted to stochastic perturbations has been investigated by means of asymptotic methods. The probability of noise-induced separatrix crossing, which drastically changes the fate of the system, is derived analytically in the case where noise is an additive Kubo-Anderson process. This theory shows how the geometry of the separatrix, as well as the noise intensity and correlation time, affect the statistics of crossing.
 Results can be applied to a wide variety of systems, and are valid in the limit where the noise correlation time scale is much smaller than the time scale of the undisturbed Hamiltonian dynamics.  
 }
\end{abstract}

\keywords{Hamiltonian system ; noise ; heteroclinic orbit}

\maketitle

\section{Introduction}
\label{intro}

 Most dynamical systems in real conditions are submitted to some noise, and 
  even weak noise can significantly affect the evolution of the system. Spectacular phenomena like noise-induced chaos \cite{frey1992,Frey1993,Simiu1996,Liu2002,Tel2008}, noise-induced synchronization \cite{Wang2009}, and noise-induced escape
 \cite{Kramers1940,Kautz1988,Beale1989,Grassberger1989,Bulsara1990,Simiu1996Melnikov}
received considerable attention in the last decades. These works allowed to clarify the role of noise and to derive relevant statistical laws. In the case of planar systems displaying homoclinic or heteroclinic orbits, it was shown that weak noise could induce separatrix crossing, and this effect was studied by means of stochastic Melnikov functions \cite{Bulsara1990,Franaszek1996,Franaszek1998,soskin2001,Khovanov2003,Khovanov2008,SuMei2008}.

  Some authors considered the effect of piecewise constant noise on dynamical systems  \cite{Anderson1954,Kubo1954,Sivathanu1995,Franaszek1998}. These perturbations are known to have non-zero correlation times, i.e. they are colored noises 
\cite{Hanggi1989}.  The advantage of this approach is that it can be applied to many realistic phenomena of interest, where some external forcing takes random  values over finite time intervals \cite{Kitahara1980,Irwin1990,Kapral1993}. 
  When noise has a bounded amplitude,  the use of stochastic Melnikov functions allows to derive a necessary condition for the appearance of chaos near homoclinic or heteroclinic cycles. These are extremely useful criteria, since they provide a threshold  below which  noise-induced chaos is impossible.  This was done by  Sivathanu {\it et al.} \cite{Sivathanu1995}, who studied a Duffing oscillator submitted to dichotomous noise, that is on-off signals that change at random times.  
  
  However, in spite of the efforts done in last decades, no general theory for the effect of such a colored noise on Hamiltonian systems has emerged so far.
 This situation corresponds to a wide class of problems involving mechanical or electrical oscillators driven by a piecewise constant forcing, or transport of inertial particles in flows, where turbulent eddies are often modeled as a random force kept constant during the lifetime of the eddy (see for example Ref. \cite{Kallio1989}).  
 The aim of the present work is to derive a general expression for the appearance of noise-induced separatrix crossing, which could be useful for systems of this kind.
  
 To achieve this goal, a weakly dissipative planar Hamiltonian system perturbed by a random force will be considered.  Because of dissipation, phase-space trajectories initiated on a heteroclinic orbit will slowly drift on one side of this orbit (also called separatrix in the following). The question investigated in this paper is twofold. Firstly, we will determine under which conditions the stochastic term can force the system to reach the opposite side of the separatrix, in spite of dissipation. Secondly, we will calculate the probability of this event. 
 We will assume that the duration of time intervals, over which the random force is constant, obeys an exponential distribution. This manifests the fact that these durations are memoryless. Our noise is therefore a Kubo-Anderson process \cite{Kubo1954,Anderson1954,Brissaud1974}, with a finite correlation time scale $\delta\tau$ which will be shown to strongly influence the probability of noise-induced crossing. The general theory is presented in section \ref{TheoGenerale}. An application is discussed in section \ref{Applic}.

 \section{Probability of separatrix crossing}

\label{TheoGenerale} 
 
 \subsection{General considerations}

 We consider a state variable ${\mathbf r}(t) = (q(t),p(t))$ evolving as
 \begin{eqnarray}
 \label{systHamil1}
 \dot q &=& \frac{\partial H}{\partial p} + \Lambda \, U_1(\mathbf r) + \varepsilon_1 \xi_1(t),\\
\dot p &=& - \frac{\partial H}{\partial q} + \Lambda \, U_2(\mathbf r) + \varepsilon_2 \xi_2(t),
\label{systHamil2}
 \end{eqnarray}
where $H(\mathbf r)$ is a differentiable scalar function, $\mathbf U = (U_1,U_2)$ is a steady vector field playing the role of a deterministic perturbation, $\Lambda$ and $\varepsilon_i$'s are
positive constants. We will assume that the deterministic perturbation is dissipative ($\nabla . \mathbf U < 0$), even though the calculations presented here are valid for any vector field $\mathbf U$.
The last terms $\varepsilon_i \xi_i$ are the components of an additive noise.
Each function $\xi_i(t)$ is a Kubo-Anderson process, as defined by Brissaud \& Frisch \cite{Brissaud1974}. It is
piecewise constant and equal to some $\xi_{ik}$ in the time interval $\tau_k < t < \tau_{k+1}$, where $\tau_k$ is an increasing series. Both the noise values $\xi_{ik}$ and the times $\tau_k$ are random variables. The former are assumed to be non-dimensional, with  zero average and  unit  variance. The latter are such that the durations $\delta \tau_k = \tau_{k+1} - \tau_k$ are memoryless and are therefore taken to have an exponential distribution with mean value $\delta \tau$ and variance $\delta\tau^2$. The number of discontinuity points $\tau_k$ within a duration $\Delta t$  obeys a Poisson distribution with parameter $  \Delta t/\delta\tau$. Under these hypotheses the autocorrelation function $\langle \xi_i(t) \xi_i(t+\Delta t) \rangle$ is equal to
$\exp(-  \Delta t/\delta\tau)$.

\begin{figure}
  \includegraphics{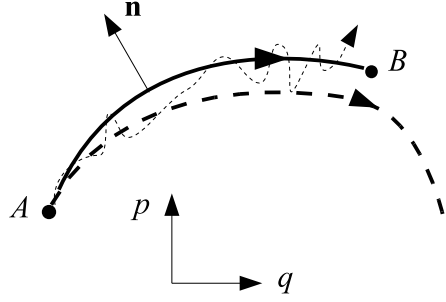}
\caption{Sketch of the undisturbed dynamics with a heteroclinic trajectory $AB$  (solid line, ($\Lambda =0$ and $\varepsilon_i = 0$)). The long-dashed line is a trajectory under weak dissipative perturbation ($\Lambda >0$ and $\varepsilon_i = 0$). The thin-dashed line is a realization of the
weakly dissipative stochastic system ($\Lambda >0$ and $\varepsilon_i > 0$).}
\label{SketchSepAB}       % Give a unique label
\end{figure}

Throughout the paper we assume that the undisturbed system (i.e. Eqs.\ (\ref{systHamil1})-(\ref{systHamil2}) with $\Lambda = 0 = \varepsilon_i$) has hyperbolic saddle stagnation points $A$ and $B$, related by a heteroclinic trajectory of equation $H(q,p) = H(A) = H(B)$  {(Fig.\ \ref{SketchSepAB})}. Such a trajectory is generally called {separatrix}, as it separates well-defined portions of the phase space. In the undisturbed dynamics, the state variable $\mathbf r(t)$ cannot cross the separatrix. In contrast, in the perturbed system, i.e. $\Lambda$ and $\varepsilon_i$'s close to zero but non-zero, crossing is possible in the sense that the sign of $H(\mathbf r(t)) - H(A)$ can vary with time.  
 This means that if $\mathbf r(t)$ travels close to $A$ at some time $t=\tau_a$, then $\mathbf r(t)$ will reach the vicinity of $B$ for some time $t=\tau_b$, but on the opposite side of separatrix $AB$. 
To follow the location of the state variable $\mathbf r(t)$ in the disturbed system, we will make use of the Hamiltonian $H$ and study the variations of $H(\mathbf r(t))$. 

Defining $\tau_a$ and $\tau_b$ as the the times when $\mathbf r(t)$ passes nearest to $A$ and $B$ respectively, the phenomenon of separatrix crossing is defined by the event $\Delta H \neq 0$, where
\begin{equation}
\Delta H \equiv H(\mathbf r(\tau_b))-H(\mathbf r(\tau_a)) = \int_{\tau_a}^{\tau_b}  
\nabla H . \dot{\mathbf r }
\, dt,
\end{equation}
and $\nabla = (     \partial / \partial q ,  \partial / \partial p )^T$.
By using Eqs.\ (\ref{systHamil1})-(\ref{systHamil2}), the variation of undisturbed Hamiltonian reads
\begin{equation}
\Delta H  = \Delta H^0 + \varepsilon_1 \int_{\tau_a}^{\tau_b} 
\frac{\partial H}{\partial q} \, \xi_1(t) \, dt
+ \varepsilon_2 \int_{\tau_a}^{\tau_b} 
\frac{\partial H}{\partial p} \, \xi_2(t) \, dt,
\label{DeltaH1}
\end{equation}
where $\Delta H^0$ is the jump of Hamiltonian in the deterministic case :
\begin{equation}
\Delta H^0 = \Lambda \int_{\tau_a}^{\tau_b} \nabla H . \mathbf U \, dt
\label{DeltaH0}
\end{equation}
which can be calculated, either analytically or numerically, as soon as $H$ and $\mathbf U$ are known. The integrals in Eq.\ (\ref{DeltaH1}) can be approximated by considering that $\mathbf r(t)$ travels close to the separatrix. 
Let $t_0$ be the time when $\mathbf r(t)$ passes nearest to $C$, the middle point of separatrix $AB$. 
Then we assume that
$\mathbf r(t) \simeq \mathbf r_0(t-t_0)$, where $\mathbf r_0(t)=(q_0(t),p_0(t))$ is a solution of the Hamiltonian system running on $AB$, that is $H(q_0(t),p_0(t)) = H(A)$ for $t \in ]-\infty,+\infty[$, with an initial position  $\mathbf r_0(0)$ equal to point $C$. 
This is a classical approximation, widely used in the calculation of Melnikov functions \cite{GH83}. Under these hypotheses, the jump of Hamiltonian can be approximated as
\begin{equation}
\Delta H  = \Delta H^0 - \varepsilon_1 \int_{\tau_a}^{\tau_b} 
\dot p_0(t-t_0) \, \xi_1(t) \, dt
+ \varepsilon_2 \int_{\tau_a}^{\tau_b} 
\dot q_0 (t-t_0) \, \xi_2(t) \, dt.
\label{DeltaH2} 
\end{equation}
Then, by making use of the fact that $\xi_i(t)$ are constant and equal to $\xi_{ik}$ over intervals $[\tau_k,\tau_{k+1}]$, with $\tau_0=\tau_a$ and $\tau_N=\tau_b$, we are led to
\begin{equation}
\Delta H  = \Delta H^0 + \sum_{k=0}^{N-1} \boldsymbol{\varepsilon}\, \boldsymbol{\xi}_k . \mathbf n(\tau_k-t_0) \, \delta s_k,
\label{DeltaHmatrice}
\end{equation}
where
\begin{equation}
\boldsymbol{\varepsilon} = \left(\begin{matrix} 
\varepsilon_1 & 0 \\
0 & \varepsilon_2 
\end{matrix}\right) ,
\end{equation}
  $\boldsymbol{\xi}_k = (\xi_{1k},\xi_{2k})^T$, and $\mathbf n$ is the unit vector perpendicular to separatrix $AB$ and pointing to the left-hand-side of $AB$. 
  In the following we write $\mathbf n_k$ in place of $\mathbf n(\tau_k-t_0)$.
  Because $\delta\tau_k$ is small compared to the typical time scale of $\mathbf r(t)$, we made use of the approximation:
\begin{equation}
\mathbf n_k  = \frac{1}{\delta s_k} \left(\begin{matrix} 
 -\delta p_{0k} \\
\delta q_{0k} 
\end{matrix}\right) 
\end{equation}
where $\delta p_{0k} = p_0(\tau_{k+1}-t_0) - p_0(\tau_k-t_0)$,   $\delta q_{0k} = q_0(\tau_{k+1}-t_0) - q_0(\tau_k-t_0)$ and 
\begin{equation}
\delta s_k = \sqrt{\delta q_{0k} ^2+\delta p_{0k} ^2} = | \dot {\mathbf r}_0(\tau_k-t_0) | \, \delta\tau_k
\end{equation}
is the discrete arc length along the path $\mathbf r_0$.
Under these hypotheses the jump of Hamiltonian reads
\begin{equation}
\Delta H = \Delta H^0 + \sum_{k=0}^{N-1} X_k
\end{equation}
where $X_k$ is the random variable
\begin{equation}
X_k = \boldsymbol{\varepsilon}\, \boldsymbol{\xi}_k . \mathbf n_k \,  | \dot {\mathbf r}_0(\tau_k-t_0) | \, \delta\tau_k.
\end{equation}

\subsection{A sufficient condition for the absence of noise-induced crossing}

   Sivathanu {\it et al.} \cite{Sivathanu1995} showed that, in the case of dynamical systems submitted to an on-off noise, there exists a threshold in the noise intensity below which noise-induced crossing cannot occur. The existence of such a threshold requires that  the noise amplitude be bounded. This means that, if the Hamiltonian drift $|\Delta H^0|$ induced by the dissipative term $\Lambda \mathbf U$ is too large, or if the $\varepsilon_i$'s are too small, noise-induced crossing will not occur. Such a sufficient condition being very useful in practice, we will generalize the results of Ref.\ \cite{Sivathanu1995}, which had been obtained for the Duffing equation, to the present problem. 
 
\begin{figure}
  \includegraphics[width=0.4\textwidth]{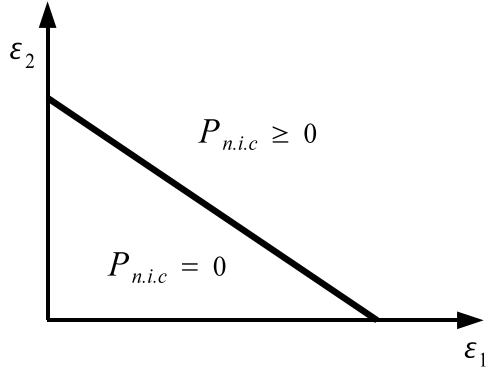}
\caption{Sketch of the plane $(\varepsilon_1,\varepsilon_2)$, together with the triangular zone where condition (\ref{Epsil12c}) is fulfilled and the system is sheltered from noise.}
\label{planEpsil}       % Give a unique label
\end{figure}

First, we assume that the noise intensities are bounded, i.e. there exists positive numbers $\xi_1^{max}$ and $\xi_2^{max}$ such that:
$$
|\xi_{1k}| \le \xi_1^{max} \quad \mbox{and} \quad |\xi_{2k}| \le \xi_2^{max}
$$
for all $k$. Then, from
Eq.\ (\ref{DeltaHmatrice}) we get:
\begin{equation}
\Big| \frac{\Delta H}{\Delta H^0}-1 \Big| \le \frac{1}{|\Delta H^0|} \sum_{i=1,2}\sum_{k=0}^{N-1} \varepsilon_i \xi_{i}^{max} |n_{ik}| u_{0k} \, \delta\tau_k,
\label{DeltaHxx}
\end{equation}
where $n_{ik} = n_i(\tau_k-t_0)$ and $u_{0k}=| \dot{\mathbf r}_0 (\tau_k-t_0) |$.  
Whatever the realization of the $\tau_k$'s, the sum over $k$ in Eq.\ (\ref{DeltaHxx}) is a Riemann sum on the interval $[\tau_a,\tau_b]$. If $\delta\tau$ is much smaller than the typical time scale of the undisturbed system, 
 this sum may be approximated by an integral:
\begin{eqnarray}
\sum_{k=0}^{N-1} \varepsilon_i \xi_{i}^{max} |n_{ik}| u_{0k} \, \delta\tau_k 
&\simeq & \varepsilon_i \xi_{i}^{max} \int_{\tau_a}^{\tau_b} |n_i(t-t_0)| u_0(t-t_0) dt\\
& \simeq & \varepsilon_i \xi_{i}^{max}
 \int_{-\infty}^{+\infty} |n_i(t')| u_0(t') dt'\\
& = &  \varepsilon_i \xi_{i}^{max} \int_{A}^{B} |n_i(s) | ds,
\end{eqnarray}
where $u_{0}(t)=| \dot{\mathbf r}_0 (t) |$ and we made use of the change of variable $t'=t-t_0$ and replaced the boundaries $[\tau_a-t_0,\tau_b-t_0]$ by $]-\infty,+\infty[$. The last equality has been obtained by introducing the curvilinear coordinate $s$ along $AB$, such that $ds = u_0(t') dt'$. 
Let $\langle . \rangle_{AB}$ denote the average over arc $AB$:
\begin{equation}
\langle f \rangle_{AB} = \frac{1}{{\cal L}_{AB}} \int_{A}^{B}  f(s)  \, ds,
\end{equation}
where ${\cal L}_{AB}$ is the length of arc $AB$.
Then, we get
\begin{equation}
 \Big| \frac{\Delta H}{\Delta H^0}-1 \Big| \le \frac{1}{|\Delta H^0|} \sum_{i=1,2}
 \varepsilon_i \xi_{i}^{max}  {\cal L}_{AB} \langle | n_i | \rangle_{AB} .
\label{DeltaHxx2}
 \end{equation}
We define noise-induced crossing as the occurrence of the event $\Delta H/\Delta H^0 < 0$, i.e. $\Delta H$ and $\Delta H^0$ have opposite signs. Indeed, as discussed in the introduction, if the dissipative term $\Lambda\mathbf U$ drives the particle $\mathbf r(t)$ on one side of the separatrix, noise-induced crossing is the fact that the particle eventually reaches the opposite side.  The probability of this event is denoted as $P_{n.i.c.}$ ("noise-induced crossing") in the following.
The inequality $\Delta H/\Delta H^0 < 0$ implies
\begin{equation}
 \Big| \frac{\Delta H}{\Delta H^0}-1 \Big| > 1
 \label{DeltaHxx3}
 \end{equation}
According to inequality (\ref{DeltaHxx2}), condition (\ref{DeltaHxx3}) cannot be fulfilled if 
\begin{equation}
 \varepsilon_1 \xi_{1}^{max}   \langle | n_1 | \rangle_{AB}
 + \varepsilon_2 \xi_{2}^{max}  \langle | n_2 | \rangle_{AB} <  \frac{|\Delta H^0|}{ {\cal L}_{AB}} .
 \label{Epsil12c}
 \end{equation}
If condition (\ref{Epsil12c}) is satisfied, noise-induced crossing will not happen and $P_{n.i.c.} = 0$. This condition corresponds to a triangular zone in the $(\varepsilon_1,\varepsilon_2)$ plane (Fig.\ \ref{planEpsil}), inside which the system is sheltered from stochastic forcing. 
Outside this zone, $P_{n.i.c.}$ can take non-zero values. This probability is calculated in the next section.

\subsection{Asymptotic expression for the jump of Hamiltonian}

We assume that noise-induced crossing occurs, and calculate the probability of this event.
Assuming that the noise intensity (that is the components of vector $\boldsymbol{\xi}_k$) and the times  $\tau_k$ are independent, we have
\begin{equation}
\langle X_k \rangle = \boldsymbol{\varepsilon}\, \langle \boldsymbol{\xi}_k \rangle . \langle \mathbf n_k \,  | \dot {\mathbf r}_0(\tau_k-t_0) | \, \delta\tau_k \rangle,
\end{equation} 
so that $\langle X_k \rangle =0$. According to the generalized central limit theorem, if $X_k$ satisfies the Lyapunov condition, i.e. if there exists $d > 0$ such that
\begin{equation}
\frac{1}{S_N^{2+d}}\sum_{k=0}^{N-1} \langle |X_k|^{2+d} \rangle \to  0 \quad \quad \mbox{as} \quad\quad N \to \infty,
\label{Lyapu}
\end{equation}
where 
$
S_N^2 = \sum_{k} \langle X_k^2 \rangle,
$
then
\begin{equation}
\frac{1}{S_N}\sum_{k=0}^{N-1} X_k \to  Z \quad \quad \mbox{as} \quad\quad N \to \infty,
\end{equation}
where $Z$ is a centred Gaussian random variable with unit variance.  
Exploiting further the fact that $\xi_{ik}$, $\tau_k$ and $\delta\tau_k$ are independent random variables we have:
\begin{equation}
S_N^2 = \sum_{k=0}^{N-1} \sum_{i=1,2}
\varepsilon_i^2  \langle \xi_{ik}^2 \rangle \, \langle n_{ik}^2
| \dot {\mathbf r}_0(\tau_k-t_0) |^2 \rangle \, \langle \, \delta\tau_k^2 \rangle,
\end{equation}
with ${\mathbf n}_k = (n_{1k},n_{2k})$. 
The variances $\langle \xi_{ik}^2 \rangle $ are equal to unity. Also, as discussed above, we assume that $\delta \tau_k$ has an exponential distribution with average $ \delta\tau$ and variance $ \delta\tau^2$. Therefore,
$\langle \delta\tau_k^2 \rangle = 2 \delta\tau^2$ and:
\begin{equation}
S_N^2 = 2\delta\tau \, \langle \sum_{k=0}^{N-1} \sum_{i=1,2}
\varepsilon_i^2  \, n_{ik}^2 
| \dot {\mathbf r}_0(\tau_k-t_0) |^2 \delta\tau \rangle .
\end{equation}
The sum in this last equation is of the form $\langle\sum_k f(\tau_k) \delta \tau\rangle$, and can be approximated as follows. We have
\begin{equation}
\sum_k f(\tau_k) \delta \tau = \sum_k f(\tau_k) \delta \tau_k - \sum_k f(\tau_k) (\delta \tau_k - \delta\tau).
\label{sommef}
\end{equation}
As $N \to \infty$, the first sum on the right-hand-side converges to an integral:
\begin{equation}
\sum_{k=0}^{N-1} f(\tau_k) \delta \tau_k \to \int_{\tau_a}^{\tau_b} f(t) dt.
\end{equation}
Indeed, whatever the realization of the $\tau_k$'s, this integral is a Riemann sum over the interval $\tau_0 = \tau_a \le t \le \tau_b = \tau_N$. 
This approximation is valid if $\delta\tau$ is much smaller than the typical time scale of the unperturbed dynamics.
The last sum in Eq.\ (\ref{sommef}) has a zero average, provided $\tau_k$ and $\delta\tau_k$ are independent:
\begin{equation}
\langle \sum_{k=0}^{N-1} f(\tau_k) (\delta \tau_k - \delta\tau) \rangle =
\sum_{k=0}^{N-1} \langle f(\tau_k) \rangle \, (\langle \delta \tau_k \rangle - \delta\tau  ) = 0.
\end{equation}
Therefore, in the limit where $N \gg 1$, we have
\begin{equation}
\langle \sum_k f(\tau_k) \delta \tau \rangle \simeq  \int_{\tau_a}^{\tau_b} f(t) dt.
\end{equation}
Finally, the sum $S_N^2$ can be approximated by
\begin{equation}
S_N^2 = 2\delta\tau \, \sum_{i=1,2} \varepsilon_i^2 \int_{\tau_a}^{\tau_b}  n_i^2(t-t_0) \, |\dot {\mathbf r}_0(t-t_0) |^2 \, dt,
\end{equation}
in the limit where $\delta\tau$ is small ($N \gg 1$).
Again, by using the change of variable $t'=t-t_0$ and replacing the boundaries $[\tau_a-t_0,\tau_b-t_0]$ by $]-\infty,+\infty[$, as usually done in the calculation of Melnikov functions \cite{GH83}, we get
\begin{equation}
S_N^2 = 2\delta\tau \, \sum_{i=1,2}\varepsilon_i^2 \int_{-\infty}^{+\infty}  n_i^2(t') \, |\dot {\mathbf r}_0(t') |^2 \, dt'
= 2\delta\tau \, \sum_{i=1,2}\varepsilon_i^2 \int_{A}^{B}  n_i^2(s) \,  { u}_0(s)  \, ds.
\end{equation}
Finally, the sum of the variances reads
\begin{equation}
S_N^2 = 2\delta\tau \,{\cal L}_{AB} \sum_{i=1,2} \varepsilon_i^2 \langle n_i^2 \, {u}_0  \rangle_{AB},
\label{SNfinal}
\end{equation}
where $u_0 = |\dot {\mathbf r}_0| = |\nabla H |$.

\subsection{Probability of noise-induced separatrix crossing}
 
The jump of Hamiltonian therefore reads, in the limit where $N \gg 1$,
\begin{equation}
\Delta H \simeq \Delta H^0 + S_N \, Z.
\label{DeltaHfinal}
\end{equation}
It is a Gaussian random variable with average $\Delta H^0$ and standard deviation $S_N$ given by Eq.\ (\ref{SNfinal}). This result will be used now to calculate the probability of noise-induced crossing.
Suppose that dissipation alone tends to drive the system towards the right-hand-side of $AB$, that is $\Delta H^0 < 0$. Then, noise-induced separatrix crossing corresponds to the event $\Delta H > 0$, i.e. noise drives the system towards the opposite side of $AB$. The probability of this event is
\begin{equation}
P_{n. i. c.} = P(Z > -\Delta H^0 / S_N)
=\int_{|\Delta H^0| / S_N}^{+\infty} g(z) \, dz
\end{equation}
where $g(z) = \exp(-z^2/2)/\sqrt{2 \pi}$. This leads to:
\begin{equation}
P_{n. i. c.}  = \frac{1}{2} \mbox{erfc}\left(  \frac{|\Delta H^0|}{2 (\delta\tau {\cal L}_{AB})^{1/2} \, (
 \varepsilon_1^2 \langle n_1^2 \, |\nabla H | \rangle_{AB} +  
 \varepsilon_2^2 \langle n_2^2 \, |\nabla H | \rangle_{AB} )^{1/2}  } 
 \right).
 \label{Pnic}
\end{equation}
which can be used whatever the sign of $\Delta H^0$. 

Equation (\ref{Pnic}) shows how the deterministic perturbation (manifested by the numerator $|\Delta H^0|$) and the stochastic perturbation (appearing at the denominator), affect the probability. In the case of isotropic noise ($\varepsilon_1=\varepsilon_2$), the denominator no longer depends on $\mathbf n$, i.e. on the shape of the separatrix. Note that the numerator $|\Delta H^0 |$ depends, in general, on the shape of $AB$.
 We also observe that the correlation time scale $\delta\tau$ has a significant effect on separatrix crossing, and that $P_{n. i. c.}$ decreases when $\delta \tau$ decreases.

\section{Application to a mechanical oscillator}
\label{Applic}

To illustrate the results of the previous section we consider the classical example of a mechanical pendulum. Indeed, oscillators  received considerable attention with and without noise (see for example San Juan {\it et al.} \cite{Sanjuan1996},  Li {\it et al.} \cite{Li2011}, 
Estevez {\it et al.} \cite{Estevez2011}), as they provide very useful test-cases.
Here, a pendulum will be used to check results (\ref{DeltaHfinal}) and (\ref{Pnic}).

\paragraph{Hamiltonian jump and probability.}
The motion equation of a weakly damped pendulum with angle $\theta(t)$ with respect to the vertical axis, when submitted to a random torque $\varepsilon \xi(t)$, where $\varepsilon > 0$ is a constant, reads
\begin{equation}
\ddot \theta = - \frac{g}{l} \sin\theta - \frac{\lambda}{m l^2} \dot \theta + \frac{\varepsilon}{m l^2} \xi(t)
\label{EqPendule}
\end{equation}
where $g$ is the gravitational acceleration, $l$ is the length of the pendulum, $m$ is its mass, and $\lambda$ is the friction coefficient. This system corresponds to Eqs.\ (\ref{systHamil1})-(\ref{systHamil2}) with $q=\theta$, $p=\dot \theta$, $\Lambda=\lambda/(m l^2)$,  $\varepsilon_1 = 0$, $\varepsilon_2=\varepsilon/(m l^2)$ and
\begin{equation}
H(q,p) = \frac{1}{2} p^2 - \omega_0^2 \cos q,
\end{equation}
where we have set $\omega_0  = \sqrt{g/l}$, and $\mathbf U(q,p) = (0, -p)^T$. In order to apply the theory developed in the previous sections, the angular frequency $\omega_0$ of the undisturbed pendulum and the noise correlation time scale $\delta\tau$
are assumed to satisfy the asymptotic condition:
\begin{equation}
\delta\tau  \ll \frac{2 \pi}{\omega_0}.
\label{dtaupetit}
\end{equation}
It will be shown below that, in this case, noise-induced crossing only appears when $\varepsilon_2$ is above some threshold $\varepsilon_2^c$. In this paragraph we assume that this condition is satisfied.

\begin{figure}
  \includegraphics[width=0.95\textwidth]{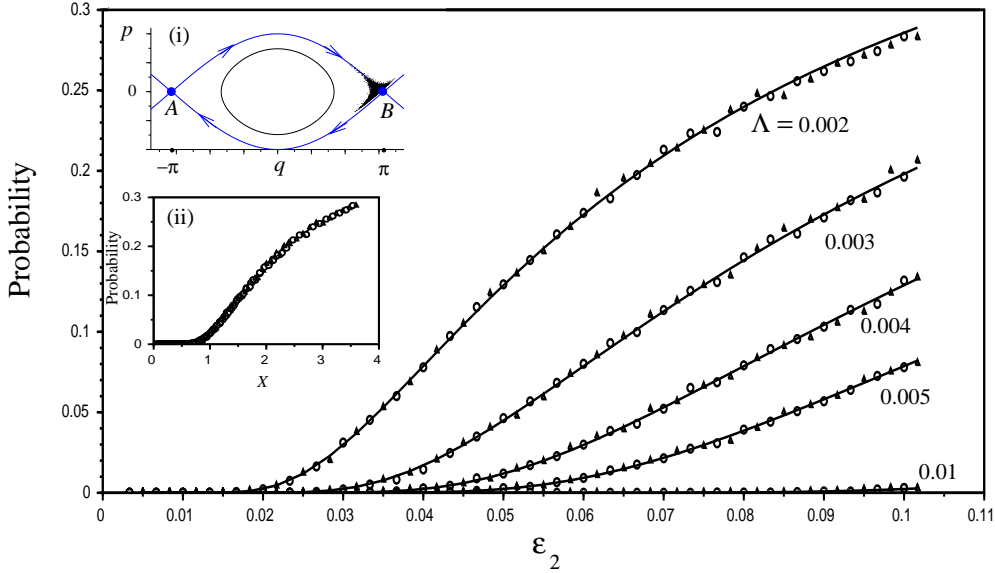}
\caption{Probability of noise-induced escape of the mechanical pendulum, for $\omega_0 = 1$ and various friction coefficients $\Lambda$. Symbols are the percentage of escape obtained from numerical simulations, by counting the number of pendulums for which $\theta > \pi$ during the run. 
Circles: uniformly distributed noise intensities. Triangles: dichotomous noise.
Solid lines: theoretical result (\ref{PnicPendule}). Inset (ii) shows the same data, but in terms of the renormalized variable  {$X=\varepsilon_2  \sqrt{\delta \tau} / \Lambda \sqrt \omega_0 $}. Inset (i) shows a typical cloud of particles in the phase space $(q,p)$, as it reaches stagnation point $B$.}
\label{ProbaPendule}       
\end{figure}

The phase portrait $H=constant$ (see inset (i) of Fig. \ref{ProbaPendule}) has the classical form of periodic cat's eyes, with hyperbolic saddle points $A=(-\pi,0)$ and $B=(\pi,0)$ related by heteroclinic trajectories. The equation of the upper
separatrix is $p_0 = 2 \omega_0 \cos(q_0/2)$. By injecting this into Eq.\ (\ref{DeltaH0}) we obtain the variation of Hamiltonian in the absence of noise:
\begin{equation}
\Delta H^0 = -\Lambda \int_{-\infty}^{\infty} p_{0}^2(t) \, dt = -\Lambda \int_{-\pi}^\pi
p_0 \, dq_0 = -8 \Lambda \omega_0.
\end{equation}
%In addition, the $p$-component of the normal vector $\mathbf n$ is:
%\begin{equation}
%n_2 = \frac{\delta q_0}{\sqrt{\delta q_0^2+\delta q_0^2}} = \frac{1}{(\omega_0^2 \sin^2(q_0/2) + 1)^{1/2}}.
%\end{equation}
%The modulus of velocity along $AB$, that is $u_0 = |\mathbf u_0 |$, satisfies
%\begin{equation}
%u_0^2 = |\nabla H|^2 =  4 \omega_0^2 \cos^2(q_0/2) (\omega_0^2 \sin^2(q_0/2) + 1).
%\end{equation}
Also, by integrating $n_2^2 u_0$ along $AB$ we get
\begin{equation}
{\cal L}_{AB} \, \langle n_2^2 u_0 \rangle_{AB} = 8 \omega_0.
\end{equation}
To ensure that the generalized central limit theorem applies, we must check that the Lyapunov condition (\ref{Lyapu}) is satisfied. Firstly, from (\ref{SNfinal}) we get:
$$
S_N^2 = K_0 \, \delta\tau,
$$
where $K_0$ is non-zero and independent of $N$ as $N \gg 1$. Therefore, since $\delta\tau = (\tau_b-\tau_a)/N$, $S_N^2$ is proportional to $1/N$ (note that this result is general). Secondly, for the pendulum we have
\begin{equation}
\langle | X_k |^{2+d} \rangle =  \varepsilon_2^{2+d} \langle \xi_{2 k}^{2+d} \rangle \langle |n_{2k} u_0 |^{2+d} \rangle \langle \delta \tau_k^{2+d} \rangle,
\label{Xk2pd}
\end{equation}
provided that the random variables $\xi_{2k}$ and $\tau_k$ and $\delta\tau_k$ are independent. For the exponential distribution of $\delta\tau_k$ considered here, and $d > 0$, we have
$$
\langle \delta \tau_k^{2+d} \rangle = \delta \tau^{2+d} \, \Gamma(d+3),
$$
where $\Gamma$ is Euler's gamma function. Also, the other terms appearing in expression (\ref{Xk2pd}) are all bounded for any $k=0,...,N-1$, so that there exists $m > 0$ independent of $N$ such that, for all $k$
\begin{equation}
 \varepsilon_2^{2+d} \langle \xi_{2 k}^{2+d} \rangle \langle |n_{2k} u_0 |^{2+d} \rangle  < m.
\end{equation}
We then have:
\begin{equation}
0 \le \frac{1}{S_N^{2+d}}\sum_{k=0}^{N-1} \langle |X_k|^{2+d} \rangle < \frac{N \, m\, \Gamma(d+3)}{K_0^{1+d/2}} \, \delta\tau^{1+d/2} = \frac{K_1}{N^{d/2}} \to 0 \quad\quad \mbox{as}\quad N \to \infty,
\end{equation}
where $K_1 > 0$ is a constant. 
The Lyapunov condition is therefore satisfied, and the Hamiltonian jump for the pendulum is the Gaussian random variable
\begin{equation}
\Delta H = -8 \Lambda \, \omega_0 + 16 \, \delta\tau \, \varepsilon_2^2 \, \omega
_0 \, Z.
\label{DeltaHPendule}
\end{equation}
The probability of noise-induced crossing for the pendulum is therefore
\begin{equation}
P_{n. i. c.}  = \frac{1}{2} \mbox{erfc}\left(  \frac{  \Lambda \sqrt {2 \omega_0} } { \varepsilon_2 \sqrt{\delta\tau}  } 
 \right), \quad \quad\mbox{if}\quad \varepsilon_2 > \varepsilon_2^c
 \label{PnicPendule}
\end{equation}
and 0 otherwise.

\begin{figure}
  \includegraphics[width=0.9\textwidth]{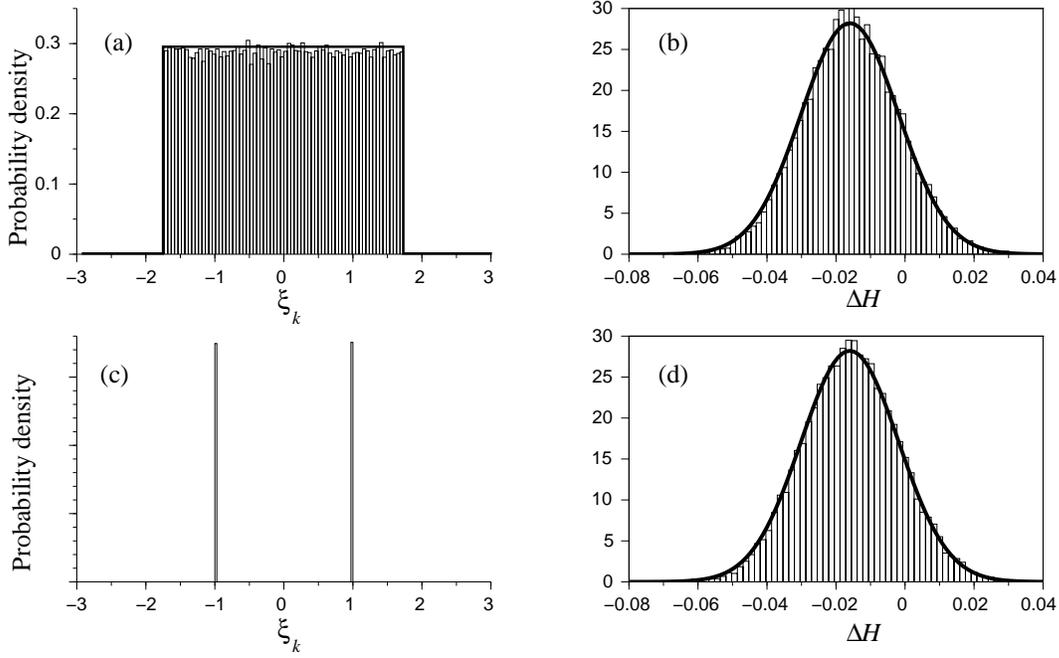}
\caption{Histograms of normalized noise amplitudes $\xi_k$ for the mechanical pendulum ((a): uniformly distributed noise, (c): dichotomous noise). Durations $\delta \tau_k$ follow an exponential distribution with average $\delta \tau = 0.005$. Graphs on the right column are the  histograms of the corresponding Hamiltonian jump ((b): uniformly distributed noise, (d): dichotomous noise). Solid lines in (b) and (d) are the normal distribution with average and variance taken from the theoretical result (\ref{DeltaHPendule}).  }
\label{histog}       
\end{figure}

\paragraph{Comparison with numerical solutions.}
Figure \ref{histog} shows the histograms of $\xi_k$  and of the resulting Hamiltonian jump $\Delta H$, obtained by solving numerically Eq.\ (\ref{EqPendule}). Graphs (a)-(b) correspond to a uniform distribution of $\xi_k$ in the range $[-\sqrt 3,+\sqrt 3]$, and (c)-(d) correspond to a dichotomous noise $\pm 1$. The parameters are  $\omega_0 = 1$, $\delta \tau = 0.005$,  $\Lambda = 0.002$, and $\varepsilon_2 = 0.05$. For both kinds of noises,  20000 runs have been conducted, with initial positions very close to $A$. They correspond to pendulums released from rest at their upper equilibrium position.
The histograms of $\Delta H$ clearly indicate that this random variable is Gaussian in both cases, with an average and a variance corresponding to the theoretical result of Eq.\ (\ref{DeltaHPendule}) (solid lines in graphs (b) and (d)).

Figure \ref{ProbaPendule} shows the theoretical  probability of Eq. (\ref{PnicPendule}) (solid line), together with probabilities obtained from numerical solutions of Eq.\ (\ref{EqPendule}) (circles and triangles).  Circles correspond to uniformly distributed noise amplitudes,  and triangles correspond to the dichotomous noise. Here, the parameters are  $\omega_0 = 1$, $\delta \tau = 0.005$, five values for $\Lambda$ have been chosen between $0.002$ and $0.01$, and $\varepsilon_2$ varies between $0$ and $0.1$. 
Circles and triangles are the percentage of pendulums which start another turn once at $B$, i.e. those for which $\theta(t) > \pi$ for some time $t$. We observe that the agreement is good, as symbols closely follow the theoretical result (\ref{PnicPendule}). As expected, when data is plotted in terms of the renormalized variable  {$X=\varepsilon_2  \sqrt{\delta \tau} / \Lambda \sqrt \omega_0 $},
all points collect along the probability   {$1/2 \, \mbox{erfc} (\sqrt 2/X)$} (inset (ii)).

\begin{figure}
  \includegraphics[width=0.7\textwidth]{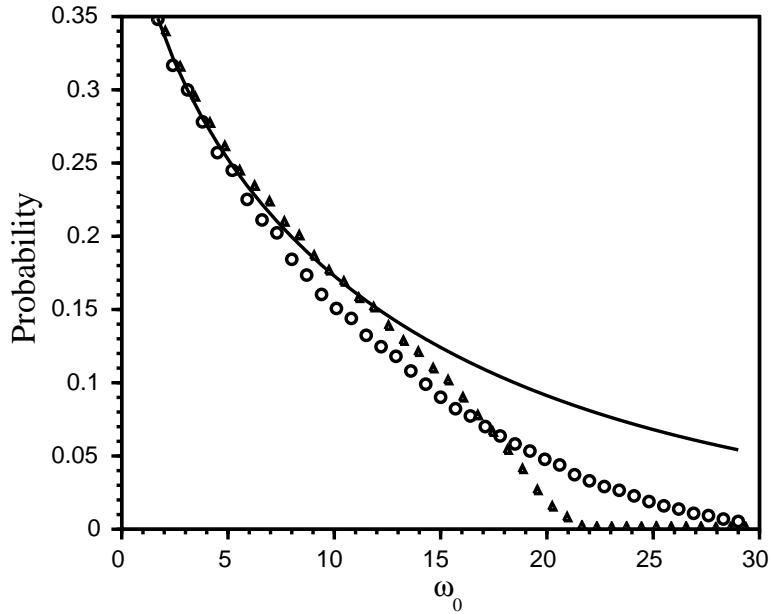}
\caption{Probability of noise-induced escape of the mechanical pendulum, for $\Lambda = 0.002$, $\varepsilon_2 = 0.06$, $\delta\tau = 0.05$, and $\omega_0$ in the range [1,30].  
Circles: uniformly distributed noise intensities. Triangles: dichotomous noise.
 For the dichotomous noise, the abrupt drop of the probability is clearly visible as $\omega_0$ approaches $\omega_{0}^c=23.6$.  }
\label{ProbaPendulew0varie}       
\end{figure}

\paragraph{Absence of noise-induced crossing for the pendulum.}  
The threshold $\varepsilon_2^c$ below which noise-induced crossing cannot occur can be readily obtained from Eq.\ (\ref{Epsil12c}). By noticing that $n_2(s) > 0$ all along the upper separatrix $AB$ and that $n_2 \, ds = dq_0$, we get ${\cal L}_{AB} \langle |n_2| \rangle = 2\pi$. The sufficient condition (\ref{Epsil12c}) leads to:
\begin{equation}
\varepsilon_2 \le \varepsilon_2^c = \frac{4}{\pi} \frac{\Lambda \omega_0}{\xi^{max} },
\label{epsi2c}
\end{equation}
where $\xi^{max} = 1$ for the dichotomous noise and $\sqrt 3$ for the uniform noise.
For the various runs of Fig.\ \ref{ProbaPendule}, $\varepsilon_2^c$ is smaller than 0.015, so that the total absence of noise-induced crossing is not visible on this figure. To observe this phenomenon, we have conducted another series of runs by choosing  $\Lambda = 0.002$, $\varepsilon_2 = 0.06$, $\delta\tau = 0.05$, and $\omega_0$ in the range [1,30]. In terms of $\omega_0$,  condition (\ref{epsi2c}) corresponds to $\omega_0 \ge \omega_0^c = \pi \varepsilon_2 \xi^{max}/ (4 \Lambda) \approx 23.6$ for the dichotomous noise and 40.8 for the uniform noise. Figure \ref{ProbaPendulew0varie} confirms that some violent phenomenon 
happens in the case of the dichotomous noise,  as the probability (triangles) drops to zero when $\omega_0$ approaches $\omega_0^c \approx 23.6$. Condition (\ref{epsi2c}) being only sufficient, the probability does not vanish at exactly $\omega_0^c$, but prior to it. As expected, the numerical probabilities shown in Fig.\ \ref{ProbaPendulew0varie} quit the theoretical law (\ref{PnicPendule})(solid line), when $\omega_0$ is above 10, since the asymptotic condition (\ref{dtaupetit}) is not strictly fulfilled there. Note that this might also affect the accuracy of the numerical estimation of $\omega_0^c$.  Concerning the uniform noise, we have checked that the circles of Fig.\ \ref{ProbaPendulew0varie} also abruptly drop to zero as $\omega_0$ approaches $\omega_0^c \approx 40.8$.

\section{Conclusion}

Noise can affect the dynamics of planar weakly dissipative Hamiltonian systems. We have studied this phenomenon in the vicinity of heteroclinic trajectories of the unperturbed Hamiltonian system, in the case where noise is piecewise constant, with a finite correlation time scale $\delta \tau$. Such a colored noise corresponds to a Kubo-Anderson process.
The analytical calculations presented here show how $\delta \tau$ affects the statistics of the Hamiltonian jump $\Delta H$ and the probability of noise-induced crossing.  Our noise having a bounded amplitude, we could also derive a sufficient condition for the impossibility of noise-induced crossing.

Our results have been illustrated by using the mechanical pendulum, submitted to weak damping and noise. Numerical simulations agree with the theoretical expressions of the Hamiltonian jump and of the probability of noise-induced crossing. A large number of complex systems can be predicted similarly, by means of formula (\ref{Pnic}). For example, it is well known that the advection of weakly inertial particles in a two-dimensional flow can be modeled by setting $(q,p) = (x,y)$, (the spatial coordinates of the particle), and $H = \psi$ (the streamfunction of the flow). The effect of particle inertia is manifested by the term $\mathbf U$, which now corresponds to the acceleration of the carrying fluid, and which has a negative divergence (see for example Refs. \cite{Maxey1987jfm,Cartwright2010}). 
Such particles are often affected by random forces, like the action of turbulent eddies, or the force due to some random electric field acting on aerosols.
If necessary, our results can be generalized to the case where the noise intensity $\varepsilon_i$ depends on the spatial position of the particle. This is a perspective of this work, which would allow to study many realistic situations 
of interest in natural or industrial dynamical systems where heteroclinic or homoclinic trajectories are present.

Finally, the phenomenon of separatrix crossing under {\it deterministic} perturbation of hamiltonian systems is well-known and has been the subject of many studies \cite{GH83,Cary1986,Neishtadt1991,Itin2002}. In practice, both deterministic and stochastic perturbations are present. The generalization of the present analysis to this case, and the derivation of a probability that would account for both effects, will be the next step of this study.

%\begin{figure*}
%  \includegraphics[width=0.75\textwidth]{example.eps}
%\end{figure*}

%\begin{table}
%\begin{tabular}{lll}
%\hline\noalign{\smallskip}
%\end{tabular}
%\end{table}

%\begin{acknowledgements}
%\end{acknowledgements}


%\bibliography{../../global}   % name your BibTeX data base

%
%% Non-BibTeX users please use
%\begin{thebibliography}{}
%%
%% and use \bibitem to create references. Consult the Instructions
%% for authors for reference list style.
%\end{thebibliography}%

\end{document}